\documentclass[12pt]{article} 
\usepackage{graphicx}
\usepackage{color}
\begin{document} 
\baselineskip 24pt
\title{A {\it p}-variable higher-order finite volume time domain  method for electromagnetic scattering problems}
\author{ A.~Chatterjee\thanks{ email: avijit@aero.iitb.ac.in, phone: +91-2225767128, fax: +91-2225722602 } \&
S.M. Joshi
\\
Department of Aerospace Engineering
\\
Indian Institute of Technology, Bombay
\\
Mumbai 400076, India}
\date{}
\maketitle
\begin{abstract}

Higher-order accurate solution to  electromagnetic scattering problems are obtained at reduced
computational cost in a {\it p}-variable finite volume time domain method. 
Spatial operators of lower, including first-order accuracy, are employed locally in substantial parts of the 
computational domain during the solution process. The use of computationally cheaper lower order spatial operators does not affect the overall higher-order
accuracy of the solution.  The order of the spatial operator
at a candidate cell during numerical simulation
can vary in space and time and is dynamically chosen  based on an order of magnitude 
comparison of scattered and incident fields at the cell centre. Numerical results are presented
for electromagnetic scattering from perfectly conducting two-dimensional scatterers subject to transverse magnetic
and transverse electric illumination.

\end{abstract}

\vspace{1.5in}

\noindent {\bf keywords}: Finite volume; Time-domain; Maxwell's equations;  Scattering; Higher-order

\newpage
\section{Introduction}
Higher-order spatially accurate representation of partial differential equations (PDE's) are used to efficiently 
resolve spatially complex physical phenomenon during numerical simulations in many fields of science and engineering. Higher-order
spatially accurate schemes are able to resolve spatial variations with lower points per wave length (PPW) in the computational
domain as compared to lower order representations. Higher-order spatially accurate methods can achieve similar accuracy levels on 
much coarser discretization compared to lower-order methods. However, higher-order spatially accurate methods tend to be more
expensive on a per-grid-point basis compared to its lower order counterparts which mitigates some of the advantages accruing
from the use of coarser meshes. Thus, there is significant motivation in developing computationally low cost higher-order methods
for numerically solving PDEs. Multigrid (MG) methods~\cite{brandt, guide}  based on cycling the numerical solution through a hierarchy of approximations 
either in space ({\it h}) or in polynomial order ({\it p}) or a combination of both have been used commonly to accelerate
convergence to steady state of boundary value problems. {\it h}-MG methods are common in both finite volume and finite element
frameworks while {\it p}-MG methods tend to be mostly restricted to mostly finite element framework~\cite{fidkowski}. 
Local {\it h} or {\it p} refinements have long been used, including for solving initial value problems,
if the length scales to be resolved are not uniform across the computational domain and can cut down significantly on
total computational time~\cite{amr, amr1}. Local refinement in the polynomial order ({\it p }) is again mostly restricted to 
finite element discretizations.  A finite volume based
solution of linear hyperbolic PDEs by cycling through successive lower order {\it p}-approximations
while retaining highest-order accuracy was proposed in Refs.~\cite{cicp, jcp}. 

In the current work we propose a {\it p}-variable finite volume framework with an emphasis on solving electromagnetic (EM) scattering
problems in the time domain. In the proposed framework, the time domain Maxwell equations which form a set of coupled linear hyperbolic PDEs,
are solved on a fixed grid but with the spatial operator formally varying in accuracy over the computational domain. The harmonic steady state solution 
obtained retains desired higher-order accuracy in spite of significant and not fixed parts of computational domain, processed using spatial operators
of lower including first-order accuracy, during the simulation. The choice of accuracy of the spatial operator, done dynamically, is based on
an order of magnitude comparison between the scattered and incident field at the cell center. The framework requires an 
unified access to spatial operators of various orders of accuracy. For the present work the ENO methodology is used to locally
obtain spatial operators of the desired accuracy but it may be possible to base it on higher-order numerical methods like spectral finite volume~\cite{svm},
ADER~\cite{ader} etc. that similarly  provides unified access to spatial operators of varying accuracy. Numerical results are presented for
electromagnetic scattering from perfectly conducting circular cylinder and airfoil.

\section{{\it p}-variable higher-order accuracy}

Consider the scalar advection equation to be the scalar representation of the
 the time-domain Maxwell's equations in differential form in a scattering process. 
The scalar advection equation is written as
\begin{equation}
\frac{\partial u}{\partial t}
+c\frac{\partial u}{\partial x}=0
\label{sca}
\end{equation}
with wave speed $ c \ge 0$.
We assume $u$ to represent a scattered field variable with
\begin{equation}
U=U_i+u
\label{tot}
\end{equation}
where $U$ and $U_i$ respectively represent the corresponding total and incident fields. All variables in equation~\ref{sca} can be nondimensionalized
as $u^{\ast}=u/U_i$, $x^{\ast}=x/\lambda$ and $t^{\ast}=t/T$.  $\lambda$ and $T$ are the wavelength and time period for the harmonic
incident wave. Further all nondimensional values $u^{\ast}$, $x^{\ast}$ and $t^{\ast}$ lie in $[0,1]$ and in terms of order of magnitude
are assumed to be $O(1)$. Equation~\ref{sca} can be written in corresponding nondimensional form as
\begin{equation}
\frac{\partial u^{\ast}}{\partial t^{\ast}}
+\frac{\partial u^{\ast}}{\partial x^{\ast}}=0.
\label{scan}
\end{equation}

The proposed {\it p}-variable method utilizes spatial operators of formally different orders of 
accuracy ($p\le m$) depending on
the order of magnitude of  scattered variables $u(x,t)$ being addressed, but always retains a local truncation error corresponding to the
the highest $m^{th}$ order accuracy. 
Spatial operators of $m$ and $(m-1)^{th}$ order formal accuracy result
in local truncation errors of similar magnitude when applied respectively to scattered variables that differ by one-order-of-magnitude.
This fact can used recursively to involve even lower order operators while retaining formal $m^{th}$ order accuracy. 
 We show this using the nondimensional form and an order-of-magnitude analysis of the local truncation error.
Discretization of the space derivative in equation~\ref{scan} with a  $m^{th}$ order accurate spatial operator results in a truncation error
with leading term given by~\cite{trunc}
\begin{equation}
a(\triangle x^{\ast}) ^m \frac{\partial^{m+1}u^{\ast}}{\partial^{m+1}x^{\ast}}
\label{tru}
\end{equation}
where $a$ is a rational number. 
In a practical finite difference type formulation approximately $10$ PPW  or more would be required for a reasonable resolution for EM scattering problems
which makes $\triangle x^{\ast}$ at least one order of magnitude less than the representative wavelength $\lambda$. Thus, $\triangle x^{\ast}
\sim O(1/10)$ in terms of order of magnitude.
Discretizing scattered variables locally of magnitude $\sim \triangle x^{\ast} \times u^{\ast}$ with a $(m-1)^{th}$ order accurate 
spatial operator will similarly lead to a truncation error with leading term 
\begin{equation}
b(\triangle x^{\ast})^{m-1} \frac{\partial^{m}\triangle x^{\ast}u^{\ast}}{\partial^{m}x^{\ast}}.
\label{tru1}
\end{equation}
In terms of order of magnitude, for constant $\Delta x^{\ast}$, 
\begin{equation}
 \frac{\partial^{m}\triangle x^{\ast} u^{\ast}}{\partial^{m}x^{\ast}} \sim
 \triangle x^{\ast} \frac{\partial^{m} u^{\ast}}{\partial^{m}x^{\ast}}
\label{ord1}
\end{equation}
using which equation~\ref{tru1} can be approximated as
\begin{equation}
b(\triangle x^{\ast})^{m} \frac{\partial^{m} u^{\ast}}{\partial^{m}x^{\ast}}.
\label{tru2}
\end{equation}
We assume
\begin{equation}
 \frac{\partial^{m} u^{\ast}}{\partial^{m}x^{\ast}}
 =\frac{\partial^{m+1} u^{\ast}}{\partial^{m+1}x^{\ast}}
=O(1)
\label{ord}
\end{equation}
since $u^{\ast}$ and $x^{\ast}$ are both $O(1)$ in the nondimensionalization process. 
This is similar to fluid mechanics boundary layer theory, where the nondimensional velocity and
distance in the streamwise direction are both $O(1)$, resulting in first and second derivatives of
the streamwise velocity in the streamwise direction also being $O(1)$~\cite{kundu}.
This further implies the leading term of the truncation error resulting from spatial operators of $m^{th}$ and $(m-1)^{th}$ 
accuracy given respectively in equation~\ref{tru} and~\ref{tru1} to be of comparable magnitude.
This can be applied recursively
to bring in spatial operators of even lower order of accuracy while locally yielding spatial accuracy comparable to the highest $m^{th}$ order accuracy.
Based on this a {\it p}-variable algorithm can be constructed to obtain inexpensively a spatially higher-order accurate steady state solution 
for a scattering process in time-domain electromagnetics or similar fields involving linear hyperbolic waves. 
The algorithm for $m^{th}$ order accuracy in a cell centered Finite Volume Time Domain (FVTD) framework can be of the form
described below and can be easily included in an existing higher-order solver,
\begin{itemize}
\item
If the cell centered scattered variable, $u(x,t) \ge  \triangle x \times U_i(x,t)$ the spatial operator is of order $m$.

\item 

For  cell centered variable $\triangle x^{n+2}\times U_i(x,t) \le u(x,t) < \triangle x^{n+1}\times U_i(x,t) $  
the spatial operator is of order $m-(n+1)$ with $n \ge 0$
\end{itemize}
with $U_i(x,t)$ assumed to be of similar order of magnitude throughout the domain and $\triangle x^n \times U_i(x,t) \sim U_i(x,t)/10^n$.

The above algorithm is used to obtain cheaply higher-order accurate solutions to the canonical problems of electromagnetic scattering
in a FVTD framework. A method of lines approach decouples the time and space discretizations and
the spatial discretization is obtained using an Essentially Non-Oscillatory (ENO) method
which allows easy access to varying orders of spatial accuracy. The current implementation 
is in the ENO-Roe form~\cite{shut,shu}, which efficiently implements
the ENO reconstruction based on the numerical fluxes instead of the cell averaged state variables and is described for the scalar law.
Equation~\ref{sca} is written as a scalar hyperbolic conservation law
\begin{equation}
u_{t}+f(u)_{x}=0,
\label{eq:shcl}
\end{equation}
has the spatial derivative at the $i^{th}$ grid point approximated as
\begin{equation}
\frac{\partial f(u)}{\partial x} |_{i}=\frac{1}{\triangle x}
(\overline{f}_{i+1/2}-\overline{f}_{i-1/2})+{\sf O}(\triangle x^{p})
\label{eq:shcl1}
\end{equation}
where ${\triangle x}$ is the grid size, $p$ the order of the scheme,
$\overline{f}_{i+1/2}$ the numerical flux function at the right cell-face.
The $r^{th}$ order accurate reconstruction of the numerical flux in the ENO
scheme is
\begin{equation}
\overline{f}_{i+1/2}=\sum_{l=0}^{r-1} \alpha_{k,l}^{r} f_{i-r+1+k+l}
\label{eq:shcl2}
\end{equation}
where $\alpha_{k,l}^{r}$ are the reconstruction coefficients and $k$
the stencil index selected among the $r$ candidate stencils. The
stencil $S_{k}$ can be written as
\begin{equation}
S_{k}=(x_{i+k-r+1}, x_{i+k-r+2},....,x_{i+k})
\label{eq:shcl3}
\end{equation}
and is locally the smoothest possible stencil.
Details regarding reconstruction coefficients and stencil selection
for ENO schemes are easily available in literature including Refs.~\cite{shut, shu}.
Extension to the multidimensional  system of equations like the time-domain Maxwell's
equations can be
obtained by decoupling the system into three scalar hyperbolic conservation
laws normal to the cell faces~\cite{cicp}.

\section{Governing Equations and Numerical Scheme}

The three-dimensional Maxwell's equations, in the differential and curl
form in free space, are expressed as
\begin{equation}
\frac{\partial {\bf B}}{\partial t}= - {\bf \nabla} \times 
\bf E 
\label{eq:curl01}
\end{equation}
\begin{equation}
\frac{\partial {\bf D}}{\partial t}={\bf \nabla} \times \bf H 
-\bf J_{i} 
\label{eq:curl}
\end{equation}
where 
${\bf B}$ is the magnetic induction,
${\bf E}$ the electric field vector,
${\bf D}$ the electric field displacement
and ${\bf H}$ the magnetic field vector.
${\bf J_{i}}$ is the impressed current density vector,
$\bf D=\varepsilon \bf E$, $\bf B=\mu \bf H$ with $\varepsilon$ and $\mu$ respectively the
permittivity and permeability in free space. 
The time-domain Maxwell's equations can also be written in a conservative 
total field form as~\cite{myong, aiaa}
\begin{equation}
\frac{ \partial \mbox{\boldmath$u$}}{\partial t}+
\frac{ \partial \mbox{\boldmath$f$}(\mbox{\boldmath$u$})}{\partial x}+ 
\frac{ \partial \mbox{\boldmath$g$}(\mbox{\boldmath$u$})}{\partial y} + 
\frac{ \partial \mbox{\boldmath$h$}(\mbox{\boldmath$u$})}{\partial z}
=\mbox{\boldmath$s$}
\label{eq:con1}
\end{equation}
where
\begin{equation}
\mbox{\boldmath$u$}\!=\!\left(\begin{array}{c}
B_{x} \\ B_{y} \\ B_{z} \\D_{x}\\D_{y}\\D_{z}
\end{array} \right), \:
\mbox{\boldmath$f$}\!=\!\left(\begin{array}{c}
0 \\  -D_{z}/\varepsilon\\ D_{y}/\varepsilon\\ 0\\B_{z}/\mu\\
-B_{y}/\mu
\end{array} \right),\:
\mbox{\boldmath$g$}\!=\!\left(\begin{array}{c}
D_{z}/\varepsilon \\ 0 \\- D_{x}/\varepsilon\\ -B_{z}/\mu\\0\\
B_{x}/\mu
\end{array} \right),\:
\mbox{\boldmath$h$}\!=\!\left(\begin{array}{c}
-D_{y}/\varepsilon \\ D_{x}/\varepsilon\\ 0\\B_{y}/\mu\\
-B_{x}/\mu\\0
\end{array} \right),\:
\mbox{\boldmath$s$}\!=\!\left(\begin{array}{c}
0 \\ 0 \\ 0\\-J_{ix}
\\-J_{y}\\-J_{iz}
\end{array} \right) 
\label{eq:con2}
\end{equation}
and subscripts indicate components in the Cartesian $x,y,z$ directions.
In two dimensions, Maxwell's equations can take two different forms corresponding to 
transverse magnetic (TM) or transverse electric (TE) waves.
The two-dimensional conservative form in general is written as
\begin{equation}
\frac{ \partial \mbox{\boldmath$u$}}{\partial t}+
\frac{ \partial \mbox{\boldmath$f$}(\mbox{\boldmath$u$})}{\partial x}+ 
\frac{ \partial \mbox{\boldmath$g$}(\mbox{\boldmath$u$})}{\partial y} 
=\mbox{\boldmath$s$}.
\label{eq:tm}
\end{equation}
The vectors in equation~(\ref{eq:tm}) for the TM waves are
\begin{equation}
\mbox{\boldmath$u$}\!=\!\left(\begin{array}{c}
B_{x} \\ B_{y} \\ D_{z}
\end{array} \right), \:
\mbox{\boldmath$f$}\!=\!\left(\begin{array}{c}
0 \\  -D_{z}/\varepsilon \\ -B_{y}/\mu
\end{array} \right),\:
\mbox{\boldmath$g$}\!=\!\left(\begin{array}{c}
D_{z}/\varepsilon \\ 0 \\ B_{x}/\mu
\end{array} \right) 
\mbox{\boldmath$s$}\!=\!\left(\begin{array}{c}
0 \\ 0 \\ -J_{iz}
\end{array} \right) 
\label{eq:tmv}
\end{equation}
while that for the TE waves are
\begin{equation}
\mbox{\boldmath$u$}\!=\!\left(\begin{array}{c}
B_{z} \\ D_{x} \\ D_{y}
\end{array} \right), \:
\mbox{\boldmath$f$}\!=\!\left(\begin{array}{c}
D_{y}/\varepsilon \\  0 \\ B_{z}/\mu
\end{array} \right),\:
\mbox{\boldmath$g$}\!=\!\left(\begin{array}{c}
-D_{x}/\varepsilon \\ -B_{z}/\mu \\ 0
\end{array} \right) 
\mbox{\boldmath$s$}\!=\!\left(\begin{array}{c}
0 \\ -J_{ix} \\ -J_{iy}
\end{array} \right) 
.
\label{eq:tev}
\end{equation}

The FVTD method solves the conservative Maxwell's equation in the integral form.
Usually a scattered field formulation is employed with the incident
field assumed to be a solution of the Maxwell's equations in free space.
Integrating the differential form of the conservation law, represented by
equation~(\ref{eq:con1}), in the absence of a source term over an arbitrary control 
volume $\Omega$
\begin{equation}
\frac{\partial \int_{\Omega} \mbox{\boldmath$u$}d\mathcal{V}
}{\partial t}
+\int_{\Omega}
\mbox{\boldmath$\nabla$}. 
(\mbox{\boldmath$F$} (\mbox{\boldmath$u$}))
 {d\mathcal{V}}=0.
\label{eq:con50}
\end{equation}
$\bf F$ is the flux vector with components $\bf f$,$\bf g$,$\bf h$ in the Cartesian $x,y,z$ 
directions with superscript `s' indicating scattered field variables.
The integral form of the conservation law to be discretized is obtained by applying the divergence theorem as
\begin{equation}
\frac{\partial \int_{\Omega} \mbox{\boldmath$u$} d\mathcal{V}
}{\partial t}
+\oint_{\mathcal{S}}\mbox{\boldmath$F$} (\mbox{\boldmath$u$})
.\mbox{\boldmath$\hat{n}$} d{\mathcal{S}}=0
\label{eq:con5}
\end{equation}
with \mbox{\boldmath$\hat{n}$} the outward unit normal vector.
The two-dimensional spatially discretized form solved for 
in a scattered and cell-centered  formulation in the present work is finally written as~\cite{myong}
\begin{equation}
A_{k}
\frac{d \mbox{\boldmath${u}$}_{k}}{dt}
+\sum_{j=1}^{4} [(
\mbox{\boldmath$\mathcal{F}$} (\mbox{\boldmath$u$})
.\mbox{\boldmath$\hat{n}$}{\mathcal{S}})_{j}]_{k}=0
\label{eq:d2d}
\end{equation}
where the numerical flux
$ [( \mbox{\boldmath$\mathcal{F}$} (\mbox{\boldmath$u$})
.\mbox{\boldmath$\hat{n}$}{\mathcal{S}})_{j}]_{k}$ approximates
 the average flux through face $j$ of cell $k$ 
and $A_{k}$ represents the area of the quadrilateral cells in structured  discretized space.
In the present work the
Maxwell's equations for TM or TE waves, in its
semi-discretized form in equation~(\ref{eq:d2d}), are solved using
higher-order ENO~\cite{shut,shu} based spatial discretization described above and
a second-order Runge-Kutta time integration.
The ENO scheme is cast in a  {\it p}-variable higher-order framework which results in highest ($m^{th}$) order accurate solutions
in the steady state, even while using spatial approximations with $p < m$ based on an order of magnitude comparison of one or more
selected field variable.
The scatterers are considered to be perfect electric conductors with
the total tangential electric field $\bf \hat{n} \times \bf E= 0$ on
the scatterer surface.
The scattered field is also assumed to be zero at the outer boundary of the computational domain where boundary conditions
are based on characteristics.

\section{Numerical Results}

\begin{figure}[ht]
\begin{center}
\includegraphics [scale=0.6]{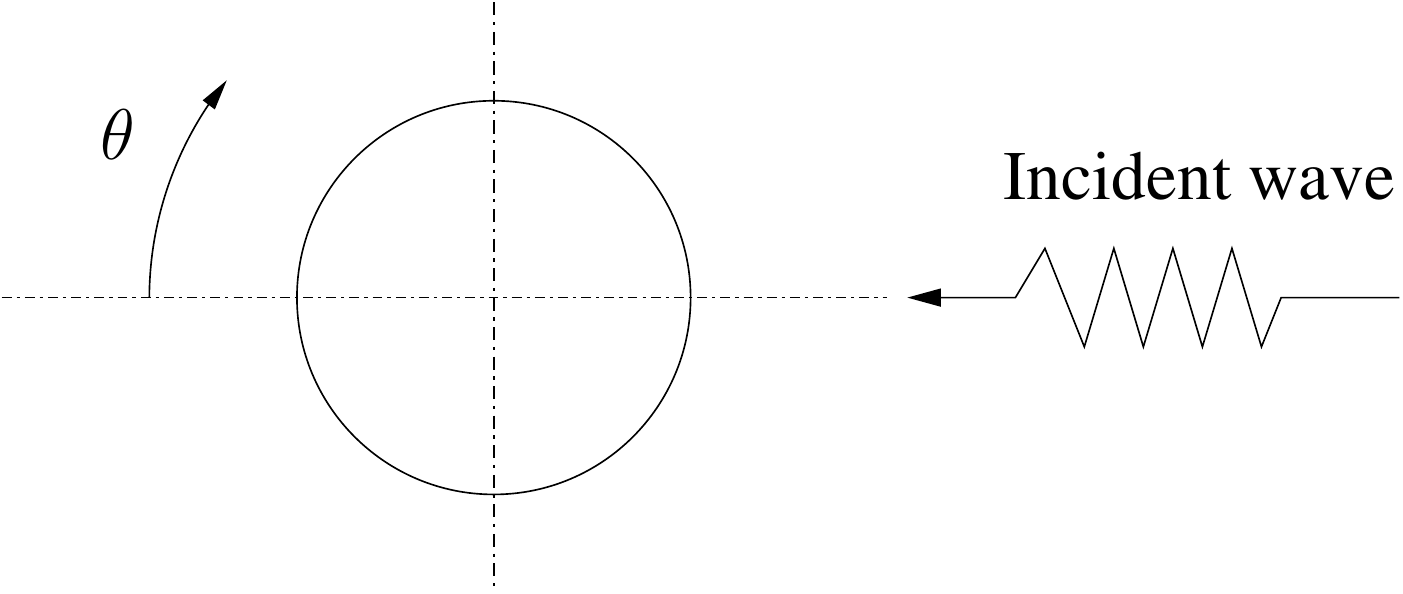}
\caption{\label{cylschem} Schematic of a circular cylinder  illuminated with an incident field }
\end{center}
\end{figure}

\begin{figure}
\begin{center}
(a)\includegraphics [scale=1.1]{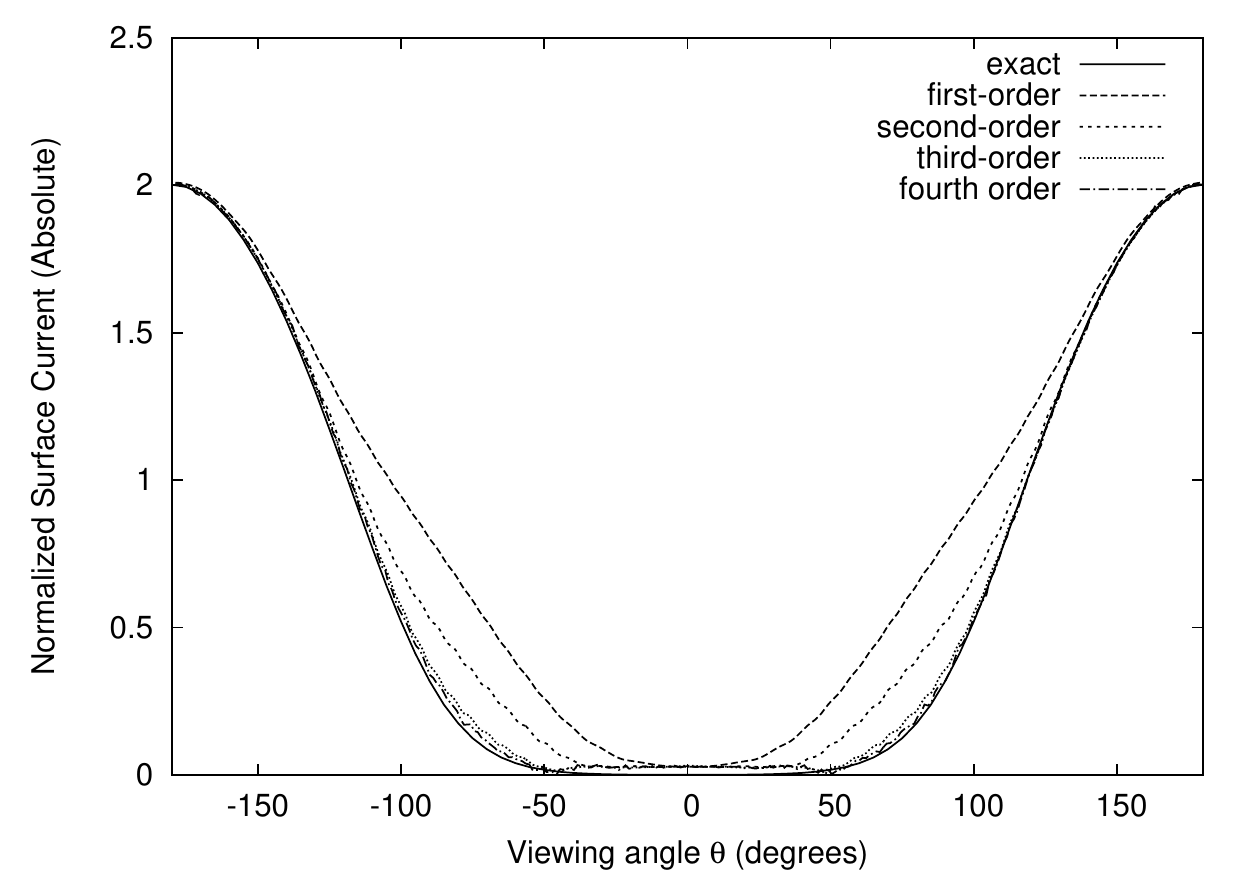}
(b)\includegraphics [scale=1.1]{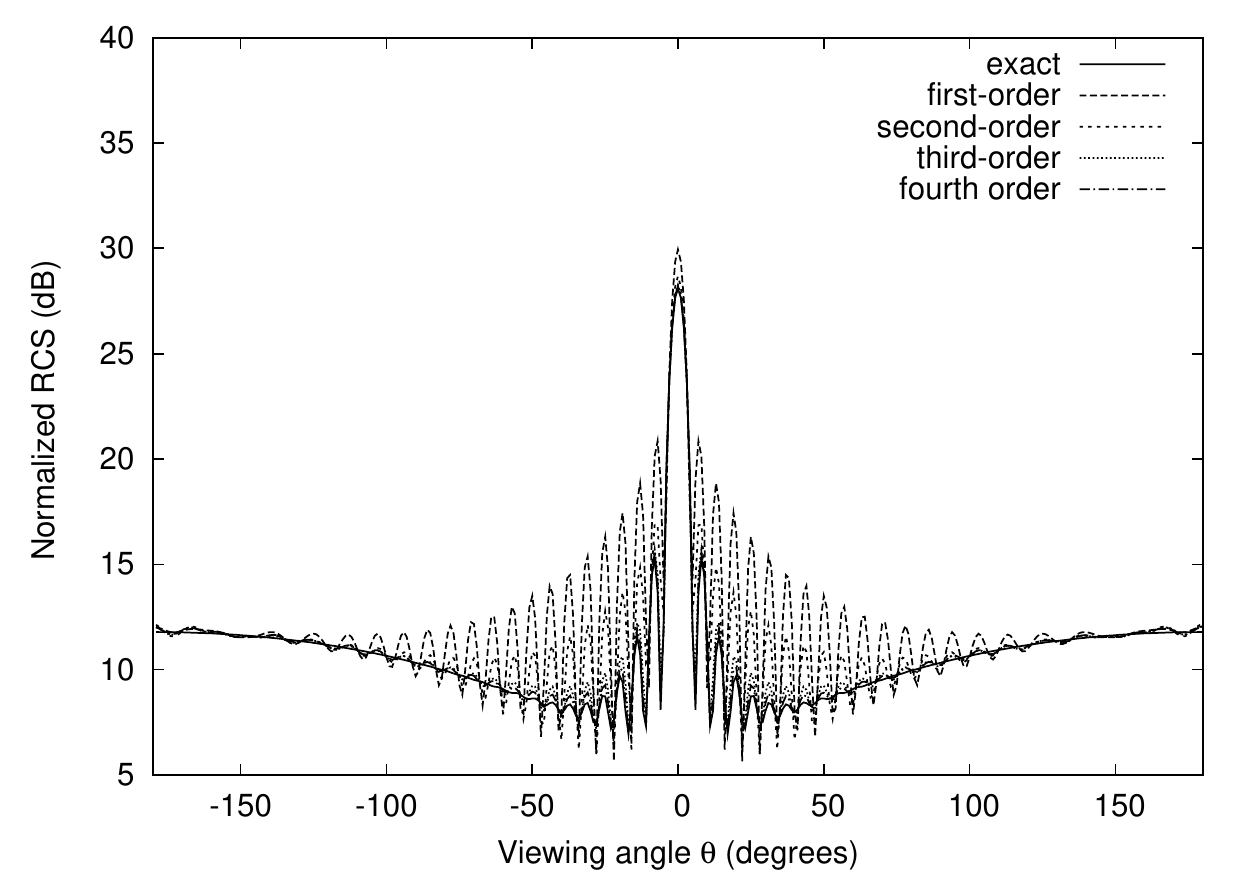}
\caption{\label{fig1} $a/\lambda= 4.8$, continuous harmonic TM illumination,
  different orders of accuracy, {\it p}-constant. (a) Surface Current Density
 (b) Bistatic RCS}
\end{center}
\end{figure}

\begin{figure}
\begin{center}
(a)\includegraphics [scale=1.1]{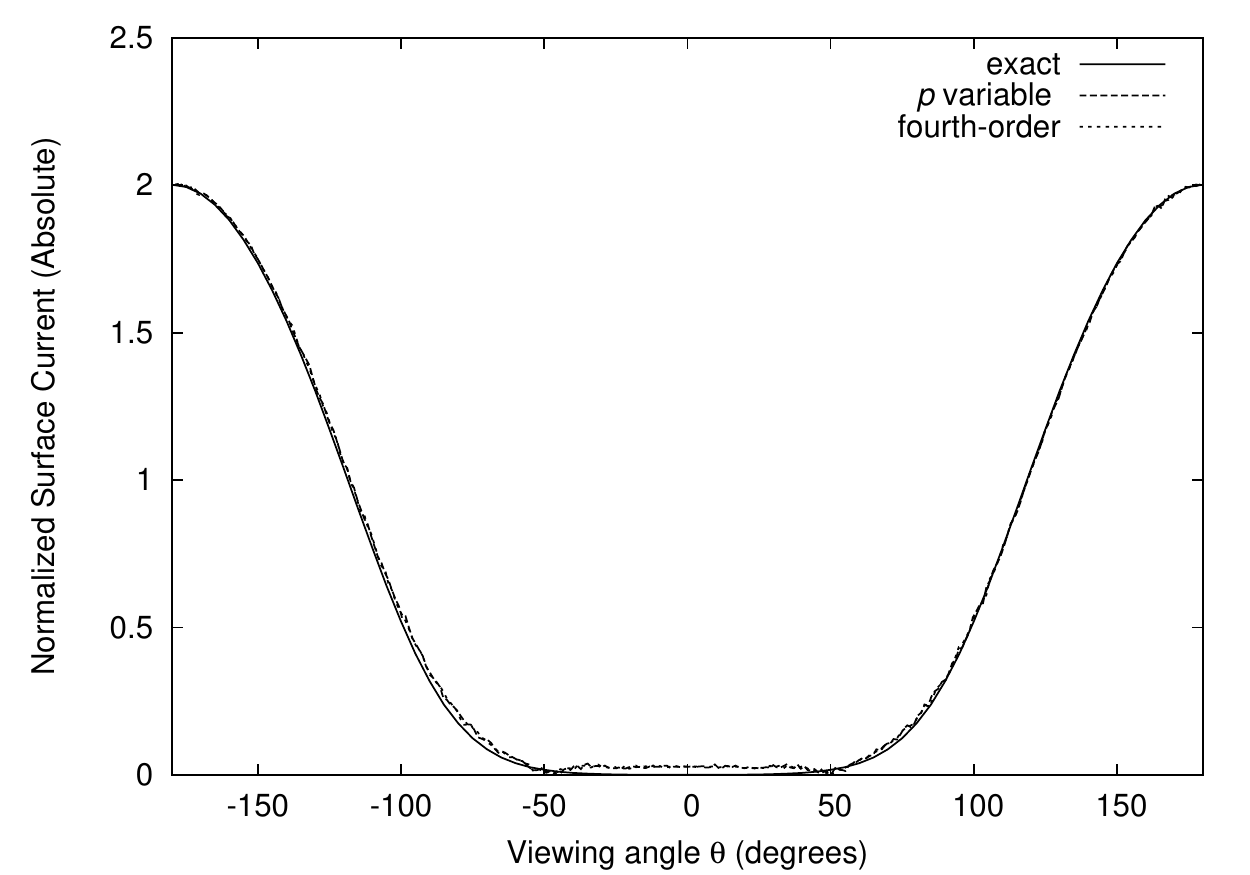}
(b)\includegraphics [scale=1.1]{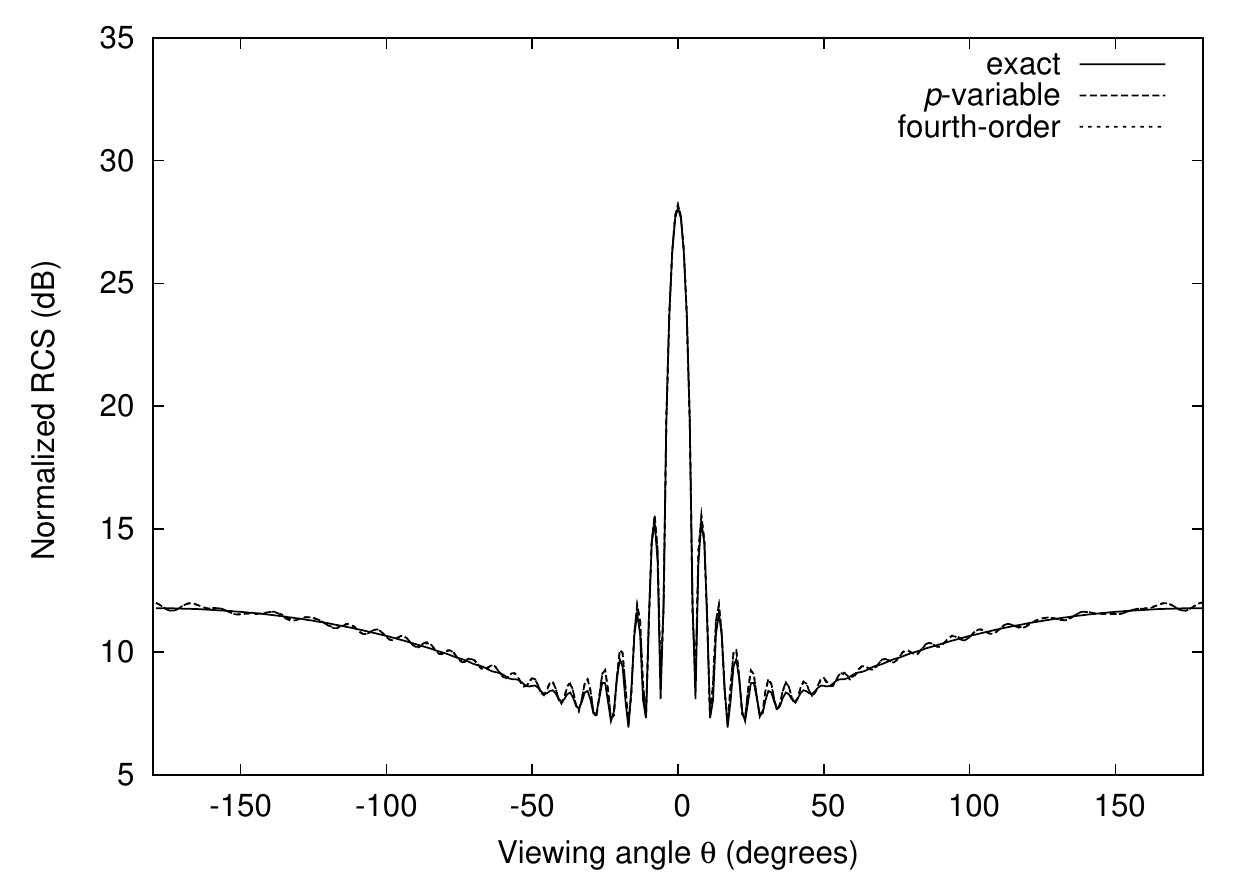}
\caption{\label{fig2} $a/\lambda= 4.8$, continuous harmonic TM illumination,
(a) Surface Current Density
 (b) Bistatic RCS;  {\it p}- variable. }
\end{center}
\end{figure}

\begin{figure}
\begin{center}
\includegraphics [scale=0.45]{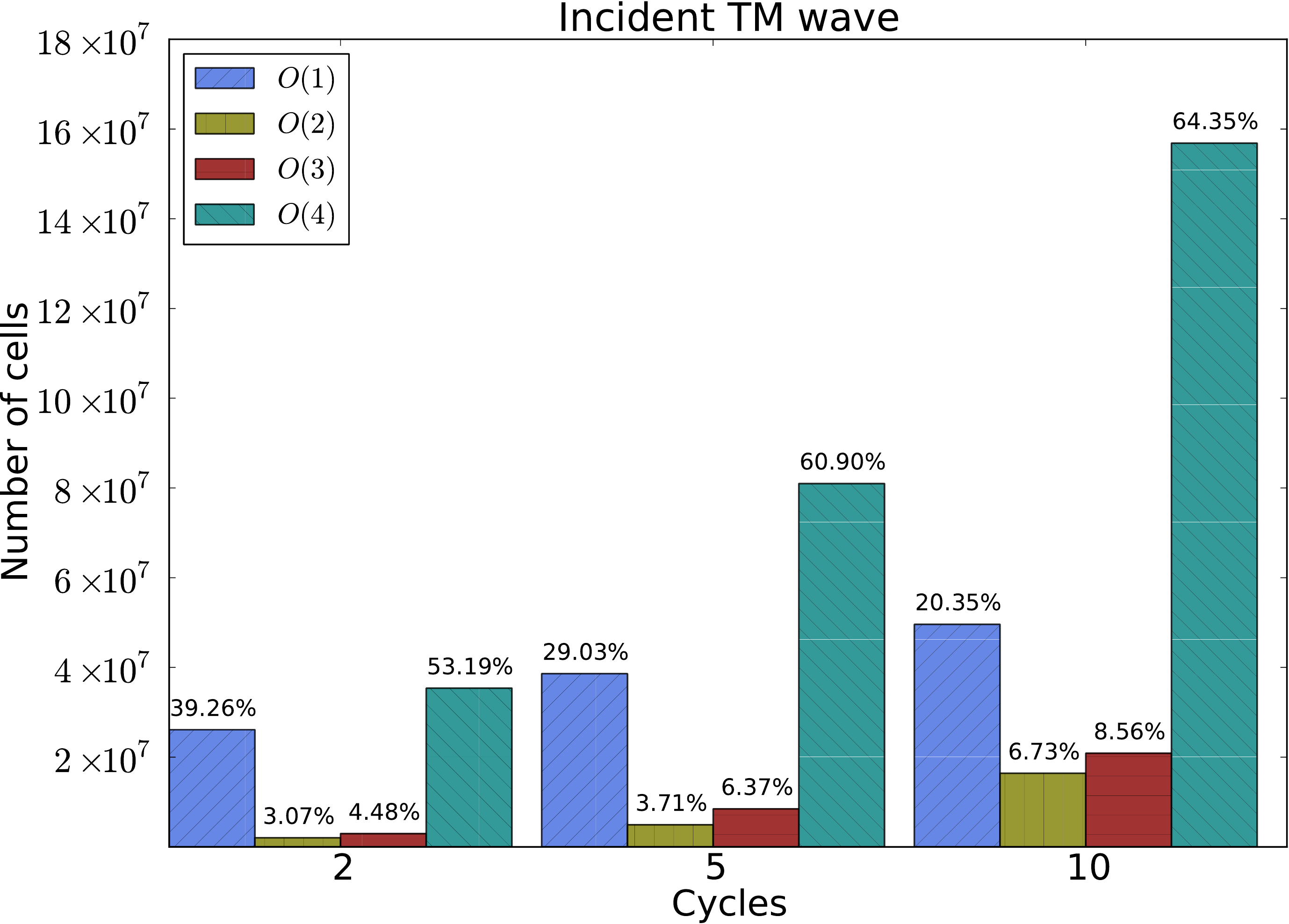}
\caption{\label{TMtable} Computational work distribution for $p$-variable method; TM case}
\end{center}
\end{figure}

Numerical results are presented for the canonical case of electromagnetic scattering from 2D perfectly conducting circular cylinders as shown in Fig.\ref{cylschem} and
compared with the exact solution. A body confirming ``O" mesh defines the computational domain with PEC boundary conditions on the 
cylinder surface and characteristic based far field conditions at the outer boundary. Results are presented for both TM and TE
continuous harmonic incident fields. Computations are performed for a fixed set of time periods of the incident harmonic wave, 
after which complex surface currents are obtained using a Fourier transform. The bistatic Radar Cross Section (RCS) 
or scattering width is then computed using a far field transformation~\cite{balanis}. 
A discussion on the number of incident wave periods to be time-stepped for attaining sinusoidal steady state
in a FDTD framework under harmonic incident excitation as attempted here is presented in Ref.~\cite{taflove}.
The first problem considered is that of the circular cylinder subject to continuous harmonic incident
TM illumination with $a/\lambda = 4.8$  where $a$ is the cylinder radius and $\lambda$ the wavelength of the incident wave~\cite{cicp, myong, deore}.
Results are shown in terms of  bistatic RCS and the absolute value of surface current 
after time stepping fixed incident time periods usually adequate for desired steady state response in such problems. Figs.~\ref{fig1}a and~\ref{fig1}b, shows sample results for a conventional implementation
for different spatial orders of accuracy on an ``O" grid with
 $300$ points in the circumferential direction corresponding to a resolution of $10$ PPW on the scatterer surface
after $5$ time periods. The number of points in the radial direction
is always kept constant at $50$. A relatively lower resolution of $10$ PPW on the cylinder surface is deliberately chosen to bring out the effect of the
numerical discretization error on the solution obtained using different spatial orders of accuracy from fourth to first. As expected, the highest 
fourth-order accurate solutions are
closest to the exact solution with first and second-order accurate solutions showing significant deviation away from near-specular-regions.
The monostatic point is located at $\pm 180^{o}$ in the bistatic plot with $0^{o}$ the perfect shadow. 
The same problem is now solved with a {\it p}-variable method with $m=4$. An order of magnitude comparison of  scattered and incident 
cellwise value of  $D_z$ is used to fix the local (cellwise)
order of accuracy $(p \le 4)$ of the spatial operator. Results are presented after $5$ time periods in 
Figures~\ref{fig2}a and~\ref{fig2}b.
and compared with exact and conventional fourth-order results. 
Results from $p$-variable method match exactly with conventional fourth-order results.
Fig.\ref{TMtable} shows the percentage 
of the computational domain over the entire simulation time processed by first, second, third and fourth-order spatial
operators while retaining an overall fourth-order accuracy. 


\begin{figure}
\begin{center}
(a)\includegraphics [scale=1.1]{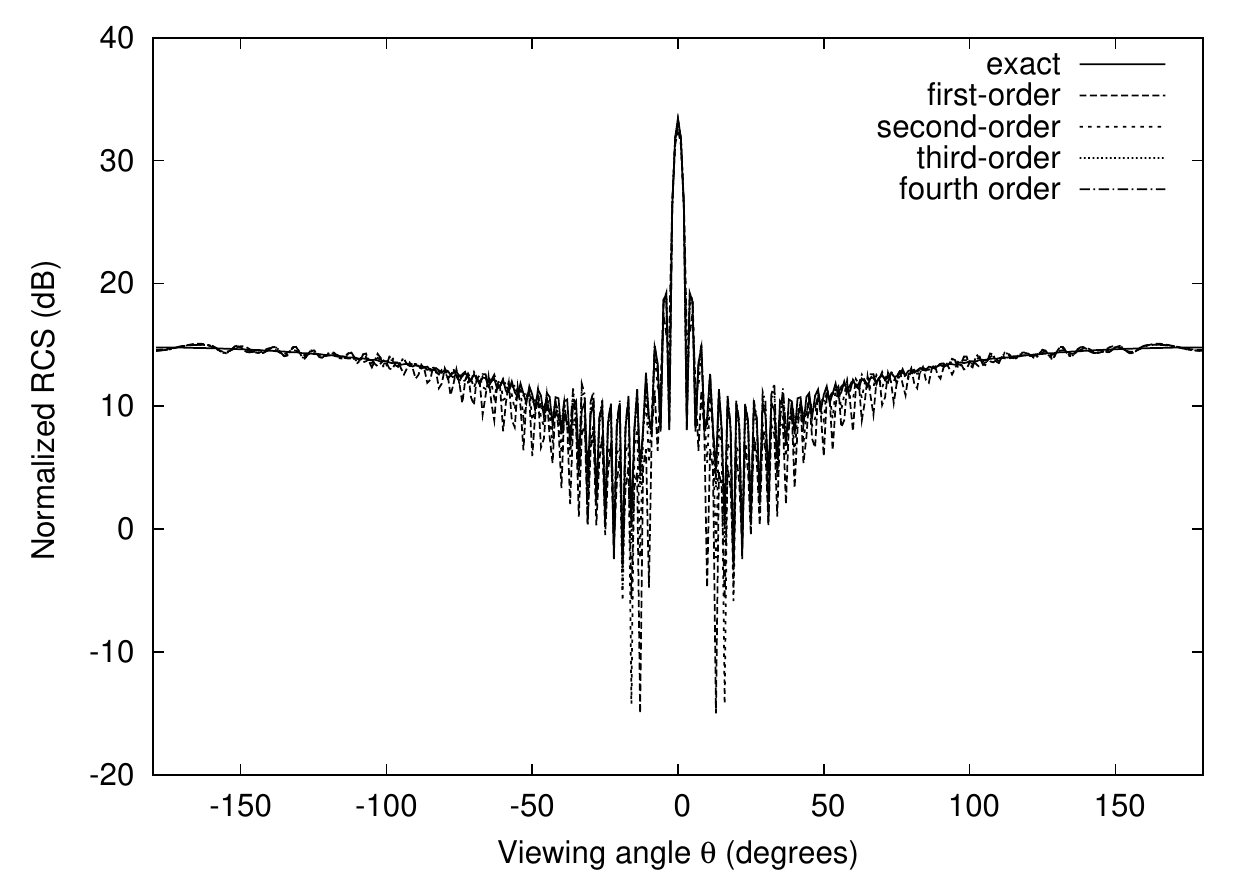}
(b)\includegraphics [scale=1.1]{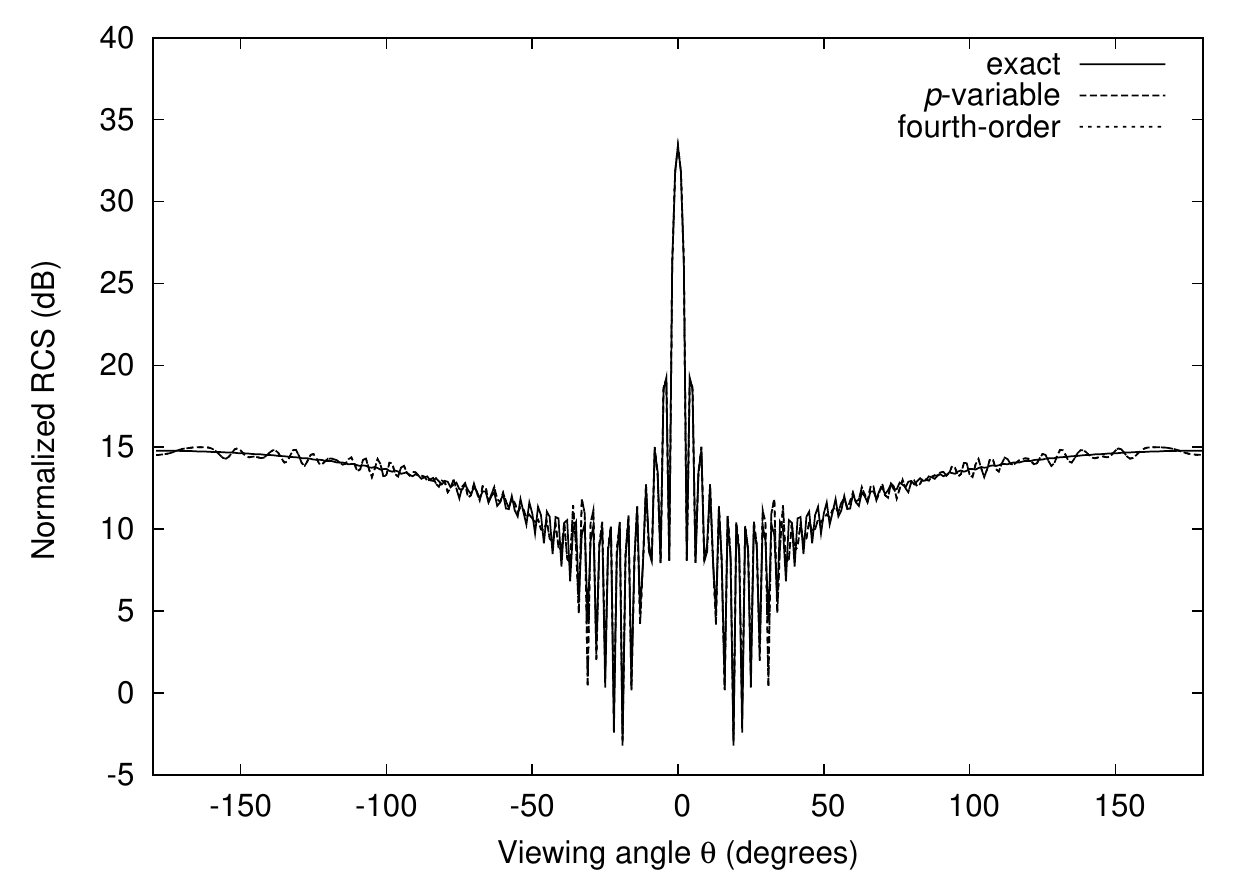}
\caption{\label{fig3} Bistatic RCS $a/\lambda= 9.6$, continuous harmonic TE illumination,
  (a) {\it p}- variable.  (b) conventional.}
\end{center}
\end{figure}

\begin{figure}
\begin{center}
\includegraphics [scale=0.45]{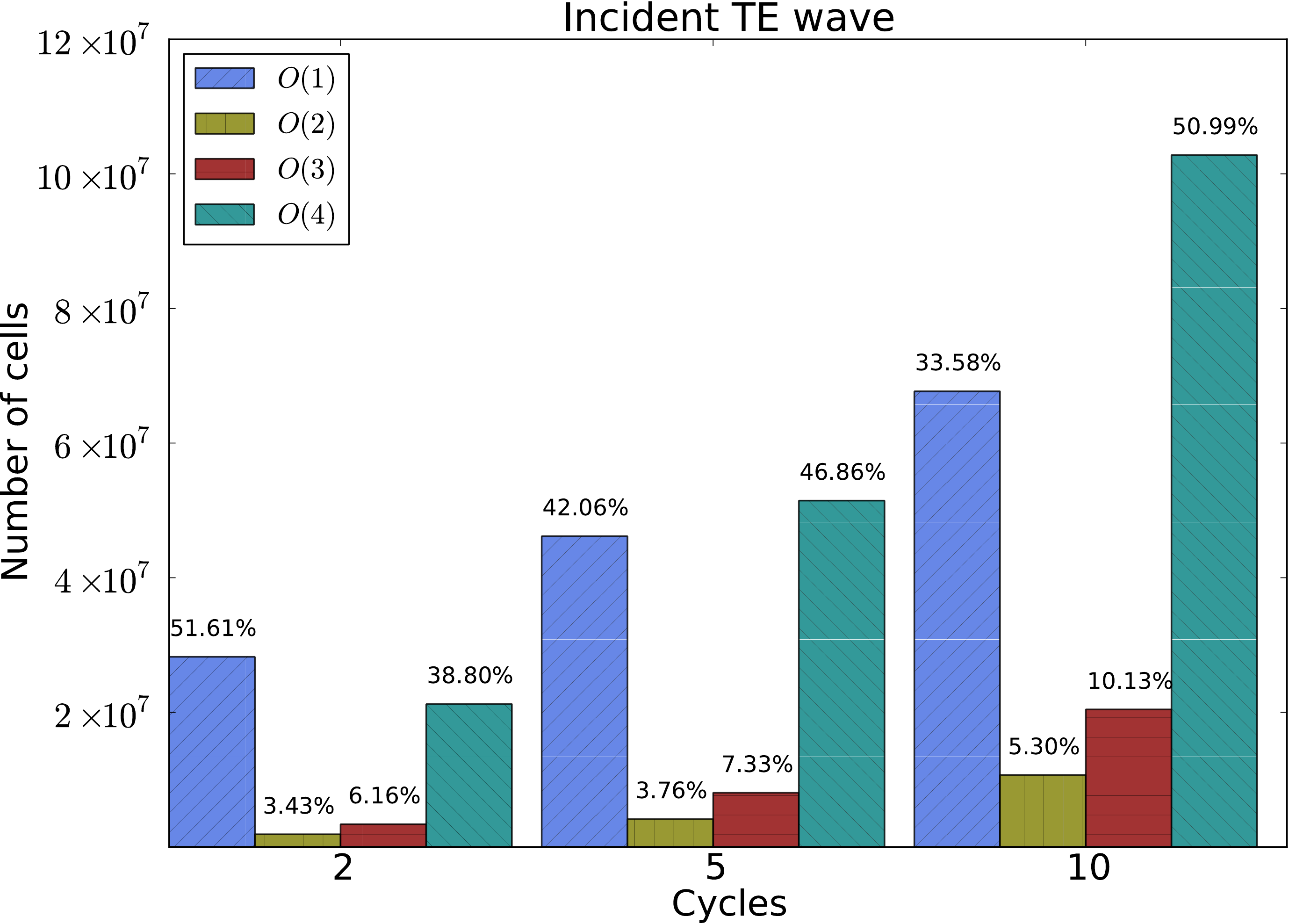}
\caption{\label{TEtable} Computational work distribution for $p$-variable method; TE case}
\end{center}
\end{figure}

The next problem considered is that of illumination by a continuous harmonic incident TE wave and $a/\lambda = 9.6$~\cite{cicp, myong, deore}. 
The ``O" grid 
with 600 points in the circumferential direction is taken so that
the resolution on the scatterer surface again corresponds to  $10$ PPW. Again, a deliberately coarse discretization is
chosen to bring out the effect of spatial order of accuracy on the obtained solution. Figure~\ref{fig3}a compares the bistatic RCS with
first, second,  third and fourth-order accuracy after $5$ time periods.  The TE solution also starts deviating from the exact solution as formal spatial order of accuracy 
goes down and this is especially apparent away from the near-specular-region.  The problem is solved with a {\it p}-variable method and $m=4$.
The choice of spatial order $p$ is based on an order of magnitude comparison of the scattered and incident value of $B_z$.
Figure~\ref{fig3}b compares the solution obtained with conventional fourth-order results. Again an almost exact match is obtained.
Fig.\ref{TEtable} lists the percentage of the computational domain processed over time by spatial operators of first, second, third and fourth-order 
accuracy while retaining formal fourth-order accuracy. 

\begin{figure}
\begin{center}
\includegraphics [scale=0.7]{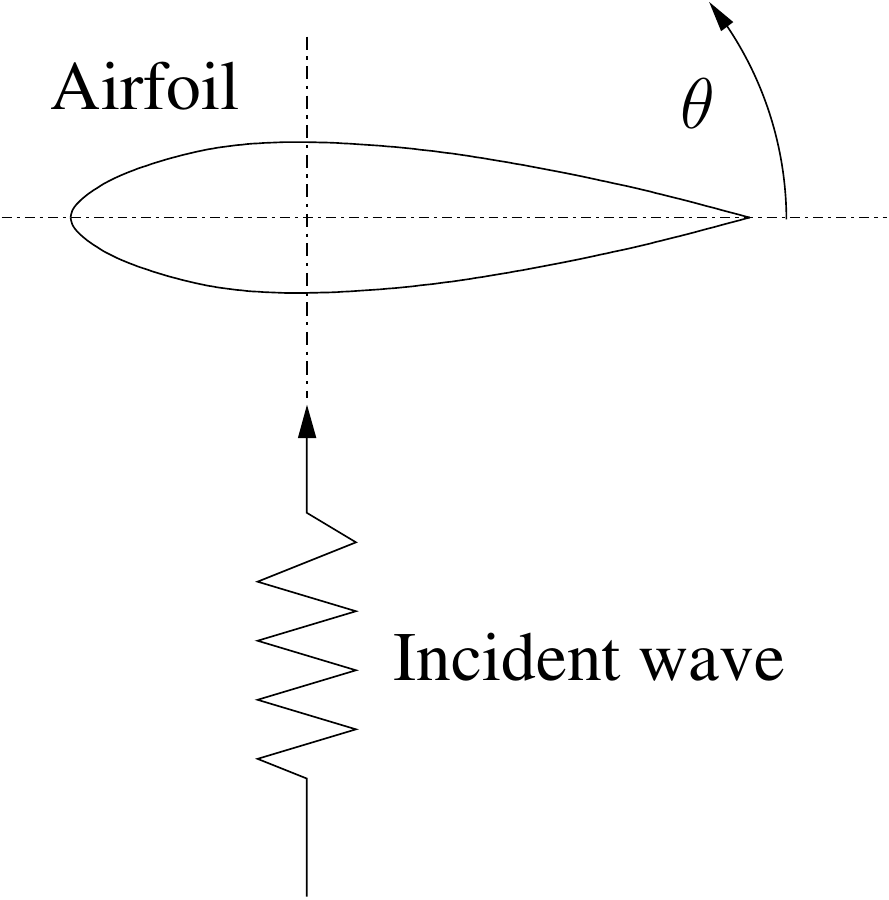}
\caption{\label{airfoil} Schematic of the NACA 0012 airfoil illuminated with an incident field }
\end{center}
\end{figure}

\begin{figure}
\begin{center}
\includegraphics [scale=1.0]{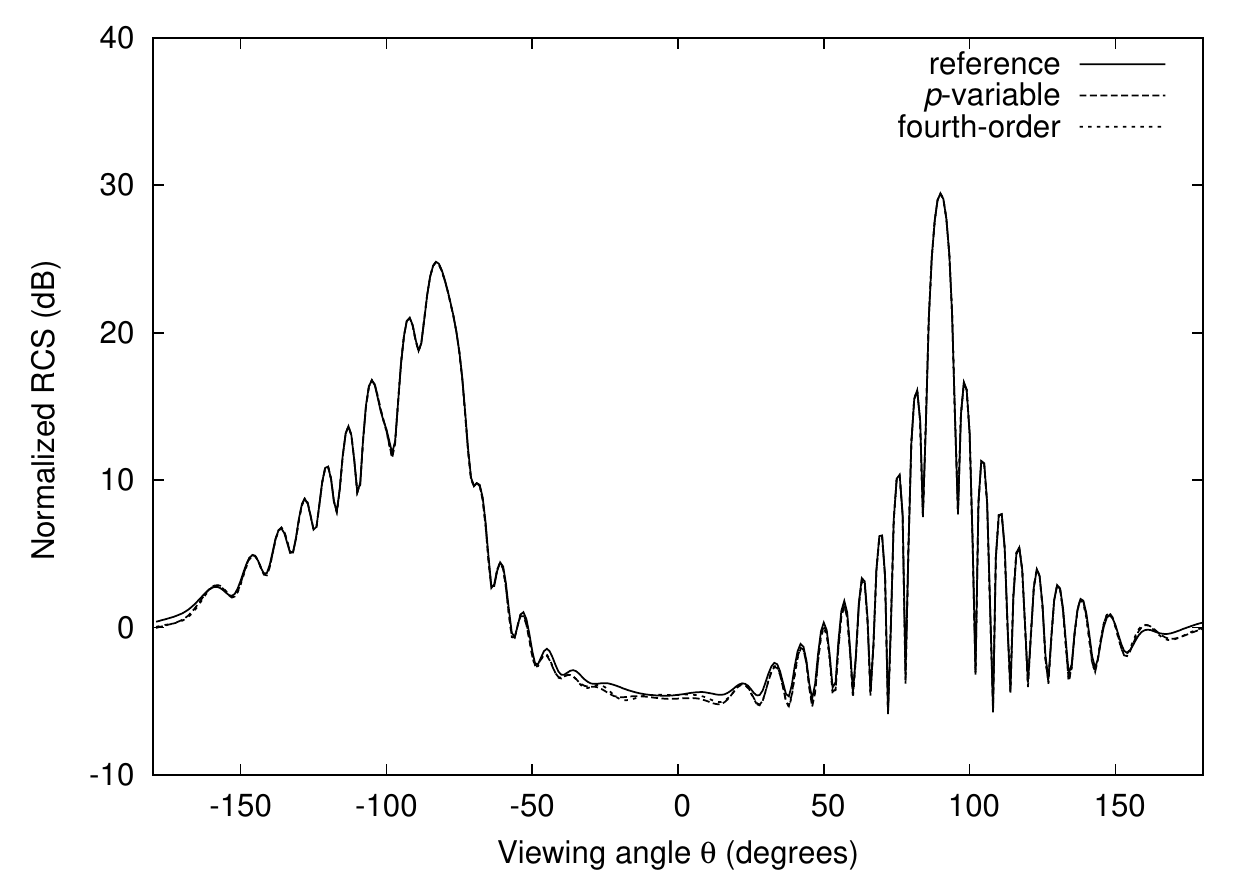}
\caption{\label{fig6} Bistatic RCS $a/\lambda= 10$, continuous harmonic TM illumination,
 {\it p}- variable and  conventional.}
\end{center}
\end{figure}

\begin{figure}
\begin{center}
\includegraphics [scale=0.42]{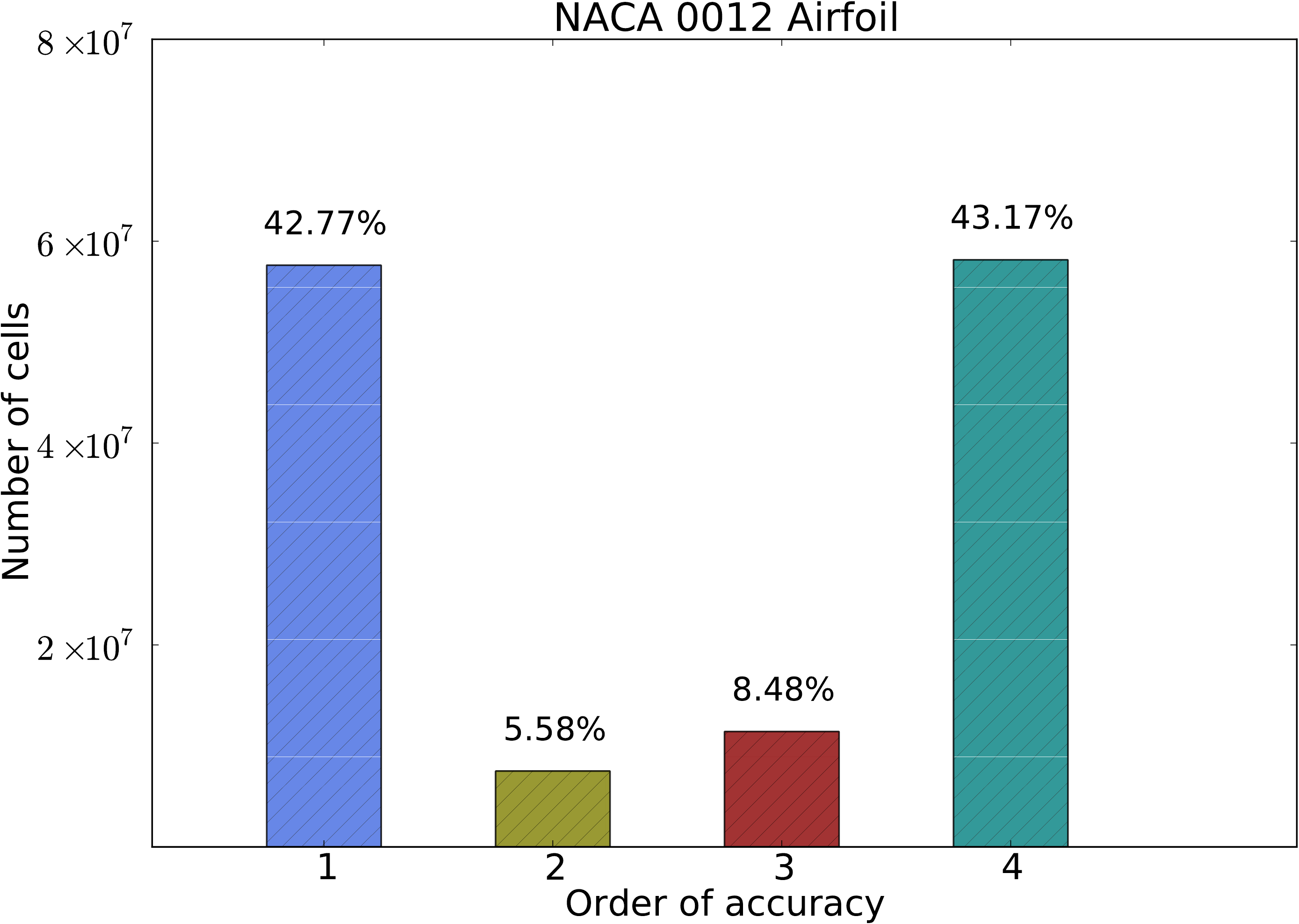}
\caption{\label{AFtable} Computational work distribution for $p$-variable method; airfoil case }
\end{center}
\end{figure}
\newpage

We also consider scattering from a perfectly conducting  NACA 0012 airfoil as shown in Fig.~\ref{airfoil}. 
The airfoil chord length is $10$ times
the wavelength of the incident harmonic TM wave at  broadside incidence~\cite{cicp, myong, deore}. Results are obtained using a 
body-fitted ``O" grid
with $200$ points around the airfoil and $50$ in the normal direction. Figure~\ref{fig6} compares RCS results after $5$ time
periods using  regular fourth-order spatial accuracy and {\it p}-variable fourth-order ($m=4$). Both results are compared with
a ``reference solution" obtained using regular fourth-order spatial accuracy but on a much finer grid with $1600$ points around the airfoil
and time stepped for $10$ time periods. Again, like in the case of the circular cylinder an almost exact match is obtained between the
conventional and {\it p}-variable method of the same formal accuracy. Fig.\ref{AFtable} lists the percentage of the computational domain 
over time processed by spatial operators $p \le 4$. The trend is similar to that for scattering from perfectly conducting circular
cylinders.

Variation in computing cost with order of accuracy for a $2$D ENO scheme is seen to follow an 
arithmetic progression \cite{cpu}.
A linear regression analysis of this data yields the computing cost per-cell at the 
$p^{th}$-order accuracy to be,
\begin{equation}
	C_p = C_1 + 3.55(p-1)
\end{equation}
where, the data is normalized with respect to the cost per-cell for a first-order accurate scheme (i.e. $C_1$).
For a $p$-variable method with $m=4$, total computing cost ($C_{total}$) can be written as,
\begin{equation}
	C_{total} = \sum_{p=1}^4 C_p n_p = (n_1 + n_2 + n_3 + n_4 ) C_1 + 3.55 n_2 + 7.1 n_3 + 10.65 n_4
\end{equation}
where, $C_p$ is the computational cost per-cell at $p^{th}$ level, and $n_p$ the total number of 
cells being processed at $p^{th}$ level.
On the other hand, the uniformly $4^{th}$-order accurate scheme will incur a cost of 
$\left(n_{T}  C_4 = n_{T}(c_1 + 3.65\times 3)\right)$ work units, where $n_T$ is the total number
of cells on the domain. 
Table \ref{costtable} shows the saving in computational cost over conventional fourth-order method in terms of work units
assuming $C_1 = 1$ unit.

\begin{table}
\begin{center}
\begin{tabular}{| c | c | c  | c | c | c | c | c | c | c |}
	
	\hline 
	\multicolumn{7}{|c|}{\bf Computational Performance - Work Units} \\
	\hline 
	{} & \multicolumn{3}{|c|}{TM Case} & \multicolumn{3}{|c|}{TE Case} \\
	\cline{2-7}  
	{  }& {Conventional} &  {$p$-variable} & {\%} & {Conventional} &  {$p$-variable} & {\%} \\

	{ Cycles }& {$O(4)$} &  {Method} & {Saving} & {$O(4)$} &  {Method} & {Saving}  \\
	\hline 
	$2$    & 7.75e08 & 4.72e08  & 39.12  & 6.37e08 & 3.11e08 & 51.14       \\
	$5$    & 1.55e09 & 1.07e09  & 30.73  & 1.28e09 & 7.29e09 & 42.96   \\
	$10$   & 2.84e09 & 2.12e09  & 25.32  & 2.35e09 & 1.48e09 & 37.01    \\
	\hline 
	\end{tabular}
	\caption{\label{costtable}Saving in computing time with $p$-variable method ($m=4$)}
\end{center}
\end{table}

\section{Conclusion}

Desired higher-order spatial accuracy can be maintained, while using lower-order spatial operators in substantial parts of the computational 
domain in a {\it p}-variable FVTD method for solving EM scattering problems. Lower-order spatial operators come at much reduced computational cost 
and can cut down  considerably on simulation time while retaining desired higher-order accuracy using the present method. An order of magnitude comparison of scattered 
and incident cell-centered EM field variables is used to decide on the local order of accuracy of the spatial operator. The local spatial
order of accuracy can vary in space and time and the proposed method can be easily integrated with existing higher-order FVTD techniques. The current implementation
uses the ENO family to access spatial operators of desired order of accuracy as dictated by the order of magnitude comparison. Results are presented
for the canonical case of EM scattering from a perfectly conducting circular cylinder as well as that of an airfoil.

\bibliographystyle{unsrt}

\end{document}